\documentclass[%
 reprint,
 amsmath,amssymb,
 aps,
]{revtex4-1}
\usepackage{graphicx}% Include figure files
\usepackage{dcolumn}% Align table columns on decimal point
\usepackage{bm}% bold math
\usepackage{amsmath,amssymb}
\usepackage{slashed}
\usepackage{graphicx}
\usepackage{dsfont}
\usepackage{subfig}
\usepackage{bm}
\usepackage[top=1.0in,bottom=1.0in,left=1.0in,right=1.0in]{geometry}
\usepackage[colorlinks,linkcolor=blue,citecolor=blue]{hyperref}
\usepackage{CJKutf8}
\usepackage{tikz-feynhand}

\begin{document}
\setlength{\oddsidemargin}{0.5cm}
\setlength{\topmargin}{-0.1cm}
\setlength{\textheight}{21cm}
\setlength{\textwidth}{15cm}
\newcommand{\be}{\begin{equation}}
\newcommand{\ee}{\end{equation}}
\newcommand{\bea}{\begin{eqnarray}}
\newcommand{\eea}{\end{eqnarray}}
\newcommand{\ba}{\begin{eqnarray}}
\newcommand{\ea}{\end{eqnarray}}

\newcommand{\fslash}{\hspace{-1.4ex}/\hspace{0.6ex} }
\newcommand{\Dslash}{D\hspace{-1.6ex}/\hspace{0.6ex} }
\newcommand{\Wslash}{W\hspace{-1.6ex}/\hspace{0.6ex} }
\newcommand{\pslash}{p\hspace{-1.ex}/\hspace{0.6ex} }
\newcommand{\kslash}{k\hspace{-1.ex}/\hspace{0.6ex} }
\newcommand{\underkslash}{{\underline k}\hspace{-1.ex}/\hspace{0.6ex} }
\newcommand{\epslash}{{\epsilon\hspace{-1.ex}/\hspace{0.6ex}}}
\newcommand{\partslash}{\partial\hspace{-1.6ex}/\hspace{0.6ex} }

\newcommand{\nn}{\nonumber}
\newcommand{\Tr}{\mbox{Tr}\;}
\newcommand{\tr}{\mbox{tr}\;}
\newcommand{\ket}[1]{\left|#1\right\rangle}
\newcommand{\bra}[1]{\left\langle#1\right|}
\newcommand{\rhoraket}[3]{\langle#1|#2|#3\rangle}
\newcommand{\brkt}[2]{\langle#1|#2\rangle}
\newcommand{\pdif}[2]{\frac{\partial #1}{\partial #2}}
\newcommand{\pndif}[3]{\frac{\partial^#1 #2}{\partial #3^#1}}
\newcommand{\pbm}[1]{\protect{\bm{#1}}}
\newcommand{\avg}[1]{\left\langle #1\right\rangle}
\newcommand{\vnabla}{\mathbf{\nabla}}
\newcommand{\notes}[1]{\fbox{\parbox{\columnwidth}{#1}}}
\newcommand{\pair}{\raisebox{-7pt}{\includegraphics[height=20pt]{pair0.pdf}}}
\newcommand{\paircrs}{\raisebox{-7pt}{\includegraphics[height=20pt]{pair0cross.pdf}}}
\newcommand{\paircc}{\raisebox{-7pt}{\includegraphics[height=20pt]{pair0cc.pdf}}}
\newcommand{\paircrscc}{\raisebox{-7pt}{\includegraphics[height=20pt]{pair0crosscc.pdf}}}
\newcommand{\pairloop}{\raisebox{-7pt}{\includegraphics[height=20pt]{pairloop.pdf}}}
\newcommand{\pairloopf}{\raisebox{-7pt}{\includegraphics[height=20pt]{pairloop4.pdf}}}
\newcommand{\pairlooph}{\raisebox{-7pt}{\includegraphics[height=20pt]{pair2looph.pdf}}}

%\preprint{APS/123-QED}

\title{Pion gravitational form factors in the QCD instanton vacuum II}

\author{Wei-Yang Liu}
\email{wei-yang.liu@stonybrook.edu}
%\affiliation{Center for Nuclear Theory, Department of Physics and Astronomy, Stony Brook University, Stony Brook, New York 11794-3800, USA}

\author{Edward Shuryak}
\email{edward.shuryak@stonybrook.edu}
%\affiliation{Center for Nuclear Theory, Department of Physics and Astronomy, Stony Brook University, Stony Brook, New York 11794-3800, USA}

\author{Ismail Zahed }
\email{ismail.zahed@stonybrook.edu}
\affiliation{Center for Nuclear Theory, Department of Physics and Astronomy, Stony Brook University, Stony Brook, New York 11794--3800, USA}

\date{\today}% It is always \today, today,
             %  but any date may be explicitly specified
\begin{abstract}
We revisit the hard QCD contributions to the pion gravitational form factors (GFFs), in
terms of the twist-2,3 pion distribution amplitudes (DA), including novel $semi$-$hard$ 
contributions from the instantons.  The pion DAs are evaluated in the QCD instanton vacuum, and then properly evolved  
%using ERBL
to higher resolution. 
The results are compared to our recent results from the QCD instanton vacuum, as well as Bethe-Salpeter calculations 
and recent lattice data. The interpolated hard and soft contributions to the pion D-form-factor, are used to derive the (gravitational) pressure and shear within the pion, with a clear delineation of their range. 
\end{abstract}

\maketitle

\section{Introduction}
\label{INTRODUCTION}

%%%%%%%%%%%%
The natural way to probe the  stress tensor (mass, energy density, pressure, sheer) structure of a hadron, would be by using
a graviton or Higgs probes. Unfortunately, none of these probes are practically available for experiments. The GFFs bring about important information on the mass and stress distribution in hadrons, which are complementary to the charge and current distributions probed by conventional electrons and photons. 

A substitute can be a 2-gluon exchange
with a hadron in the elastic regime. 
%may well be the substitute needed for understanding the hadron 
Recently, the gluon gravitational form factors (GFFs)  of the nucleon have been extracted at JLab by the GlueX collaboration~\cite{Ali:2019lzf,GlueX:2023pev} and the SOLID collaboration~\cite{Duran:2022xag}, using near-threshold charmonium photoproduction.
The results from these  experiments have initiated a number of theoretical and lattice studies, all aimed at determining the GFFs of various lowest hadrons.

On more general ground, the GFFs are accessible through generalized parton distributions (GPDs), through deeply virtual photon or meson production~\cite{Diehl:2003ny,Belitsky:2005qn,Mezrag:2022pqk} (and references therein). Current and future 
experiments both at the JLab facility,
and in the upcoming electron ion collider (EIC)~\cite{AbdulKhalek:2021gbh,Anderle:2021wcy}, are set to provide more empirical results and insights to the  quark and gluon composition of the hadronic GFFs. 
A number of lattice analyses of the GFFs have now been carried out in~\cite{Hackett:2023nkr,wang2024trace},
with ab initio insights to these issues, with models
for their interpretation~\cite{Broniowski_2008,Frederico_2009,Masjuan_2013,Fanelli_2016,Freese_2019,Krutov_2021,Raya:2021zrz,Xu:2023izo,Li_2024,Broniowski:2024oyk}.

At asymptotically large momentum transfer, the hadronic FFs can be evaluated using perturbative QCD, thanks to asymptotic freedom and factorisation)~\cite{Callan:1969uq,Gross:1973id,Politzer:1973fx,Collins:1989gx}. 
The following ``QCD counting rules" apply to the hadronic GFFs as well. Recently, the  leading twist-2,3 pion contribution, were partially evaluated in~\cite{Shuryak:2020ktq}, with the inclusion of the instanton semi-hard contribution. The full leading twist-2 contribution was evaluated in~\cite{Tong:2021ctu,Tong:2022zax}. In this paper, we will revisit and extend 
these calculations to include the full twist-3 contributions, which are also important in the semi-hard regime. The factorisation  analysis follows that used in standard hard exclusive processes at large momentum transfer, with the results expressed as hard kernels folded with twist-2,3 pion distribution amplitudes.

Most semi-inclusive and exclusive processes involve momenta in the few GeV$^2$ regime. This domain of momenta transfers is commonly known  as a $semi$-$hard$ regime. The possibility of accessing bottomonia at the future EIC will put this exchange into larger,   10 GeV$^2$, range. As
two of us have argued \cite{Shuryak:2020ktq} that QCD factorisation does not hold in the semi-hard range, and suggested to supplement it with non-perturbative contributions. The aim of this paper is to detail these suggestions for the pion GFFs. 

Central to our analyses of the pion distribution amplitudes, is the QCD vacuum model knkown as the ``instanton liquid"~\cite{Schafer:1996wv} (and references therein).
Among other phenomena, it explains how the QCD vacuum breaks conformal and chiral symmetries. The quark condensate and effective quark masses provide
a mechanism at the basis of  (most) hadronic masses. The 
exact nature of the gauge fields at the origin of these breakings is still studied and debated,  but there is clear and increasing  evidence how much the QCD
vacuum is populated by instantons and anti-instantons (pseudoparticles), representing tunneling paths
between classical vacuaa with different topological charges. Various ``cooling" (noise removal) techniques have revealed
details of the tunneling landscape~\cite{Leinweber:1999cw}.  These
pseudoparticles are very effective at breaking spontaneously chiral symmetry through fermionic zero modes with fixed chirality (left or right). They do explain many parameters
of the pions specifically, as we will show below in detail.

The paper is organized as follows: In section~\ref{SECII} we review the  twist 2,3 pion DAs discussed in~\cite{Shuryak:2020ktq,Kock:2021spt} in the context of the ILM. We also show how they evolve under the renormalization group evolution. In section~\ref{SECIII} we detail the pQCD (hard) contributions to the pion gravitational form factors up to twist-3. The analysis allows the separation 
of the quark and gluon contributions, for the pion EMT as well as the pion A,D-form factors. The semi-hard contribution following chiefly from the instanton non-zero-modes  is also discussed. In section~\ref{SECIV} we show how the results from
the soft regime we derived recently~\cite{Liu:2024jno}, are matched to the semi-hard and
hard regimes here, to account for the pion EMT and A,D-form factors through all kinematical range. Our  results are in good agreement with the recently reported 
lattice results in the soft regime, and predict their extrapolation to the hard regime. In section~\ref{SECV} we use our results to analyze the pion radial pressure
and shear content, with comparison to recent results using the Bethe-Salpeter construction. Our results show how the pressure and shear are budgeted among the hard and soft contributions in the pion. Our conclusions are in section~\ref{SECVI}.

\section{Pion distribution amplitudes}
\label{SECII}
The semi-hard and hard contributions to the pion  form factors, were initially  analyzed by two of us in~\cite{Shuryak:2020ktq} for all probes, with a partial analysis of the gravitational form factor. The full analysis of the latter up-to twist-2 was subsequently discussed in~\cite{Tong:2021ctu,Tong:2022zax}. 
Thanks to factorization in QCD, a hard graviton probing the pion receives contributions from a hard kernel times pertinent pion DAs.
In brief, the analysis of most form factors in~\cite{Shuryak:2020ktq} made use   of the pion DA up-to twist-3 defined by
 \begin{widetext}
 \begin{eqnarray}
\label{WF1}
&&\int_{-\infty}^{+\infty}\frac{p^+dz^-}{2\pi}e^{ixp^+z^-}\left<0\left|\overline{d}_\beta(0)W[0,z]u_\alpha(z)\right|\pi^+(p)\right>\nonumber\\
&&=
\bigg( \frac{iF_\pi}4\gamma^5\bigg(\slashed{p}\,\varphi^A_{\pi}(x)
-\chi_\pi \varphi_{\pi}^P(x)+i \chi_\pi
\sigma_{\mu\nu}\frac{p^\mu p^{\prime\nu}}{p\cdot p^\prime}  \frac{\varphi^{T\prime}_{\pi}(x)}6\bigg)\bigg)_{\alpha\beta}
\end{eqnarray}
or more explicitly
\begin{eqnarray}
\label{Pion_DA}
&&\varphi^A_{\pi}(x)=
\frac  1{iF_\pi}\int_{-\infty} ^{+\infty} \frac{dz^-}{2\pi}e^{ixp\cdot z}\left<0\left|\overline{d}(0)\gamma^+\gamma_5W[0,z]u(z)\right|\pi^+(p)\right>\nonumber\\
&&\varphi^P_{\pi}(x)=
\frac  {p^+}{F_\pi\chi_\pi}\int_{-\infty} ^{+\infty}  \frac{dz^-}{2\pi}e^{ixp\cdot z}\left<0\left|\overline{d}(0)i\gamma_5W[0,z]u(z)\right|\pi^+(p)\right>\nonumber\\
&&\frac{\varphi^{T\prime}_{\pi}(x)}6=
\frac  1{F_\pi\chi_{\pi}}\frac {p^\mu p^{\prime \nu}p^+}{p\cdot p^\prime}\int _{-\infty} ^{+\infty} \frac{dz^-}{2\pi}e^{ixp\cdot z}\left<0\left|\overline{d}(0)\sigma_{\mu\nu}\gamma_5W[0,z]u(z)\right|\pi^+(p)\right>
\end{eqnarray}
\end{widetext}
Here $$W[0,z]=\exp\left(-ig\int_0^{z^-}dz' A^+(z')\right)$$ refers to the Wilson line in the light cone direction. The pion weak decay constant is $F_\pi=\sqrt 2 f_\pi\approx 133\,\rm MeV$, and the pion pseudoscalar strength $\chi_\pi$ is defined as
 \begin{equation} 
\label{CHI}
\chi_\pi=\frac{m_\pi^2}{(m_u+m_d)}
\end{equation}

The evaluation of the pion DAs (\ref{Pion_DA}) have been considered in \cite{Kock:2021spt}, using the instanton liquid model (ILM) of the QCD vacuum,
with the key parameters~\cite{Shuryak:1981ff}
%. In brief, two key parameters of it  are  the vacuum instanton density and size, respectively.
\begin{equation}
\label{PIA}
n_{I+A}\equiv \frac 1{R^4}\approx \frac 1{ {\rm fm}^{4}} \qquad\qquad\frac{\bar \rho}R \approx  \frac 13  
\end{equation}
These parameters amounts to a small instanton packing fraction 
$$\kappa\equiv \pi^2\bar\rho^4 n_{I+A}\approx 0.1$$
The hadronic scale $R=1\,{\rm fm}$ emerges as  the mean quantum tunneling rate of the pseudoparticles. 
All the  DA's in (\ref{Pion_DA}) are normalized to $1$ at the scale $\mu=1/\rho\simeq631$ MeV. 
Here the value of $\chi_\pi (\mu=1/\rho)$ is chosen to be $0.551$ GeV to make sure the twist-3 DA's are properly normalized. \cite{Shuryak:2020ktq}

The explicit form of the pion DAs in the ILM read~\cite{Kock:2021spt}
%The use of  the pion LFWFs \eqref{PWF} into \eqref{Pion_DA}, yields 
\begin{widetext}
\begin{eqnarray}
&&\varphi^A_{\pi}(x)=
\frac{2N_c M^2}{f^2_\pi}\frac{1}{8\pi^2}\theta(x\bar{x})\int_0^{\infty} dk^2_\perp  \frac{1}{x\bar{x}m^2_\pi-k_\perp^2-M^2} \mathcal{F}^2\bigg(\frac{k_\perp}{\lambda^A_\pi \sqrt{x\bar x}}\bigg)\\
&&\varphi^P_{\pi}(x)=
\frac{N_c M}{f^2_\pi\chi_\pi}\frac{1}{8\pi^2}\frac{\theta(x\bar{x})}{x\bar{x}}\int_0^{\infty} dk^2_\perp  \frac{k_\perp^2+M^2}{x\bar{x}m^2_\pi-k_\perp^2-M^2} \mathcal{F}\bigg(\frac{k_\perp}{\lambda^P_\pi \sqrt{x\bar x}}\bigg)\\
&&\frac{\varphi^{T\prime}_{\pi}(x)}6=
\frac{N_c M}{f^2_\pi\chi_\pi} \frac{1}{8\pi^2}\frac{x-\bar{x}}{x\bar{x}}\theta(x\bar{x})\int_0^{\infty} dk^2_\perp  \frac{k_\perp^2+M^2}{x\bar{x}m^2_\pi-k_\perp^2-M^2} \mathcal{F}\bigg(\frac{k_\perp}{\lambda^T_\pi \sqrt{x\bar x}}\bigg)
\end{eqnarray}
\end{widetext}
where $\bar x =1-x$, $\theta(x\bar{x})=1$ in the physical domain $0<x<1$.
Note that we have corrected the numerator in the integrand of $\varphi_\pi^P$
in comparison to~\cite{Kock:2021spt}. Here $\mathcal{F}$ is the so called instantonic form factor, which is the transverse profile of the  fermion zero modes in the ILM
 \bea
 {\cal F}(k)=\big(z(I_0(z)K_0(z)-I_1(z)K_1(z))'\big)^2|_{z=\frac 12 k\rho}\nonumber\\
 \eea
Here the normalization constant $C_\pi=-\sqrt{N_c}\ M/f_\pi$ by the Goldberger-Treiman relation \cite{Liu:2023yuj,Liu:2023fpj}. The parameters $\lambda_\pi$ are chosen to make sure the distribution amplitudes are properly normalized \cite{Kock:2021spt} at $\mu=1/\rho$.
\begin{align}
\label{lambda_pi}
    \lambda_\pi^A&=2.464 & \lambda_\pi^P&=1.070 & \lambda_\pi^T&=1.076
\end{align}

The pseudoscalar and tensor DAs are not independent but  tied by the current identity \cite{Shuryak:2020ktq}
\bea
    &&\partial_{\nu}\bar{d}(0)\sigma^{\mu\nu}\gamma^5u(z)=\nonumber\\
    &&-\partial^\mu\bar{d}(0)i\gamma^5u(z)+m\bar{d}(0)\gamma^{\mu}\gamma^5u(z)
\eea
therefore they share the same coupling strength  $\chi_\pi$. Its value is  fixed by the divergence of the axial-vector current, and the partially conserved axial current (PCAC) relation
\begin{widetext}
\begin{equation}
    (m_u+m_d)\langle0|\bar{d}(0)i\gamma^5u(0)|\pi^+(p)\rangle=- (m_u+m_d)\mathrm{tr}\left[i\gamma^5\left(\frac{iF_\pi}{4}\gamma^5\chi_\pi\right)\right]\int_0^1dx\varphi^P_\pi(x)=- (m_u+m_d)\chi_\pi F_\pi
\end{equation}
\end{widetext}

The twist-2 pion DA asymptotes   $\varphi^A_{\pi^+}(x)\rightarrow 6x\bar{x}$ at high resolution. No anomalous dimension appears in it, due to the axial current conservation. However, the anomalous dimensions are non-zero in the cases of the twist-3 pion DAs. Therefore, their normalization constant  varies with changing resolution. The twist-3 DAs asymptote  
\bea
\varphi^P_{\pi^+}(x,\mu)\rightarrow && 1\times\left(\frac{\alpha_s(\mu)}{\alpha_s(\mu_0)}\right)^{-\frac{3C_F}{b}}\nonumber\\
\varphi^T_{\pi^+}(x,\mu)\rightarrow && 6x\bar{x}\left(\frac{\alpha_s(\mu)}{\alpha_s(\mu_0)}\right)^{\frac{C_F}{b}}
\eea
owing to their conformal collinear spin and nonzero anomalous dimension. The anomalous dimension of $\varphi^P_{\pi}$ is the same as the running quark mass \cite{Tarasov:1982plg}.

\begin{figure}
\subfloat[\label{fig:DAA}]{%
\includegraphics[height=4cm,width=1.\linewidth]{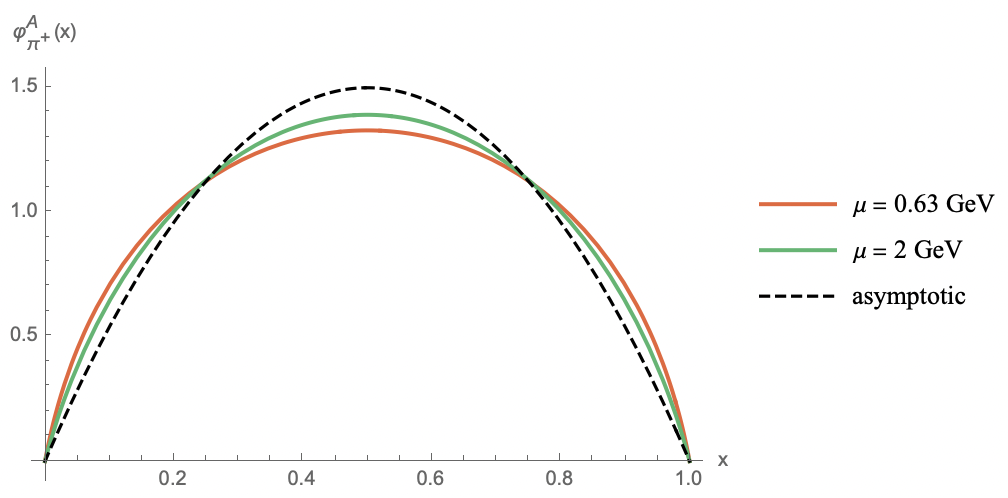}%
}\hfill
\subfloat[\label{fig:DAP}]{%
\includegraphics[height=4cm,width=1.\linewidth]{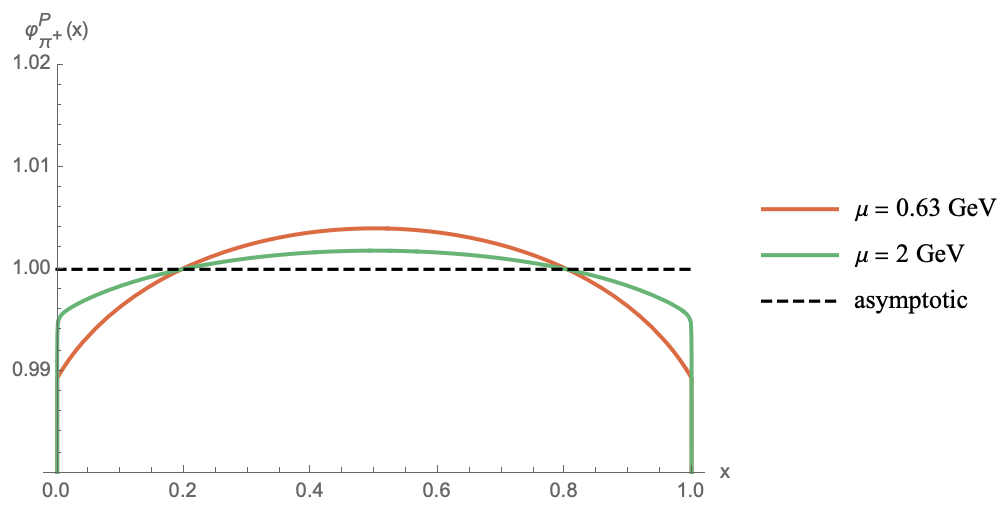}%
}\hfill
\subfloat[\label{fig:DAT}]{%
\includegraphics[height=4cm,width=1.\linewidth]{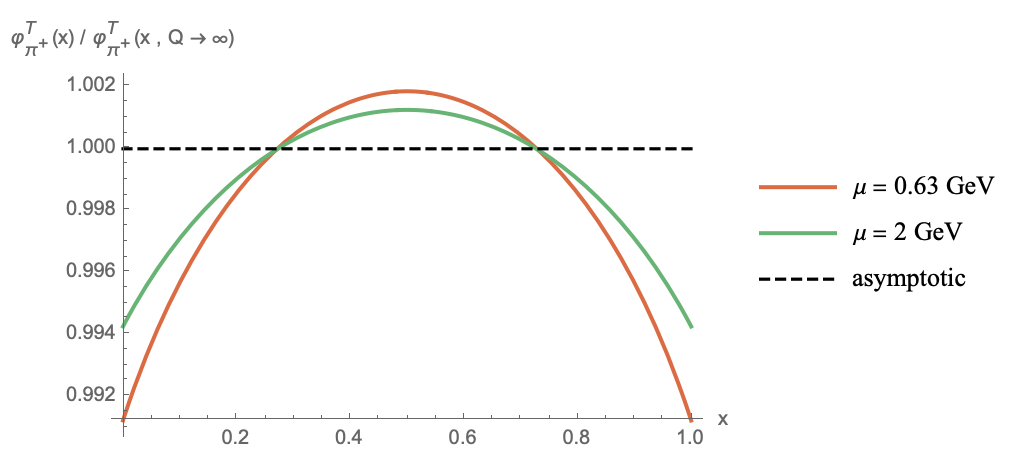}%
}
\caption{Pion twist-2,3 DAs. All DAs 
are normalized to 1 at the low resolution point $\mu=\mu_0$,
fixed by the inverse instanton size.}
\label{fig:pion_DA}
\end{figure}

%\fbox{suggestion: add a plot of ratio of components in the last case}

\subsection{The DA evolution}
The DA's with  asymptote $6x\bar{x}$ are expanded in terms of Gegenbauer polynomials $C^{3/2}_n(x-\bar{x})$, while the one that asymptotes $1$ is expanded in terms of Gegenbauer polynomials $C^{1/2}_n(x-\bar{x})$,
as expected from perturbative QCD in the Bjorken limit. More specifically,  we have 
\begin{equation}
\varphi^{A,T}(x,\mu)=6x\bar{x}\sum_{n=0}^\infty C^{3/2}_n(x-\bar{x}) a^{A,T}_n(\mu) 
\end{equation}
where due to the orthogonality of the Gegenbauer polynomials, 
\begin{equation}
\begin{aligned}
    &a^{A,T}_n(\mu)=\\
    &\frac{2(2n+3)}{3(n+1)(n+2)}\int_0^1dyC^{3/2}_n(y-\bar{y})\varphi^{A,V,T}(y,\mu)
\end{aligned}
\end{equation}
Similarly, we have~\cite{Shifman:1980dk}
\begin{equation}
\varphi^{P}(x,\mu)=\sum_{n=0}^\infty C^{1/2}_n(x-\bar{x}) a^{P}_n(\mu)  
\end{equation}
where due to the orthogonality of the Gegenbauer polynomials, 
\begin{equation}
    a^{P}_n(\mu)=(2n+1)\int_0^1dyC^{1/2}_n(y-\bar{y})\varphi^{P}(y,\mu)
\end{equation}

The DA evolution is given by
 Efremov-Radyushkin-Brodsky-Lepage (ERBL) procedure  in terms of the Gegenbauer polynomials,
\begin{equation}
a^{A,P,T}_n(\mu)=a^{A,P,T}_n(\mu_0)\left(\frac{\alpha_s(\mu)}{\alpha_s(\mu_0)}\right)^{\frac{1}{b}\gamma^{A,P,T}_n}
\end{equation}
with the anomalous dimensions \cite{Braun:2016wnx,Shifman:1980dk} 
\bea
    \gamma^A_n&=&C_F\left[-3+4\sum_{j=1}^{n+1}\frac{1}{j}-\frac{2}{(n+1)(n+2)}\right]\nonumber\\
\gamma^P_n&=&C_F\left[-3+4\sum_{j=1}^{n+1}\frac{1}{j}-\frac{8}{(n+1)(n+2)}\right]\nonumber\\
\gamma^T_n&=&C_F\left[-3+4\sum_{j=1}^{n+1}\frac{1}{j}\right]
\eea
and 
\bea
C_F&=&\frac{N_c^2-1}{2N_c}\nonumber\\
\alpha_s(\mu)&=&\frac{4\pi}{b\ln\left(\frac{\mu^2}{\Lambda_{QCD}^2}\right)}\nonumber\\
b&=&\frac{11}{3}N_c-\frac{2}{3}N_f
\eea
and $\Lambda_{QCD}=250$ MeV. Here $C^m_n(z)$ are Gegenbauer polynomials.

All DAs are normalized to 1 at the low resolution point fixed by the inverse instanton size $\mu=\mu_0=1/\rho\approx 0.63\, \rm GeV$. In Fig.~\ref{fig:DAA} we show our results for $\varphi_\pi^A(x)$ at low resolution at $\mu=0.63\,\rm GeV$(solid-red), after evolution at $\mu\,\rm GeV$ (solid-green) and the asymptotic limit (dashed-black). The analogous results for $\varphi_\pi^{P}(x)$ 
with the same color coding is shown in Fig.~\ref{fig:DAP}. Our results for the
ratio of the tensor DA $\varphi_\pi^{T}(x)$ at low and high resolution, are shown in Fig.~\ref{fig:DAT} also with the same color coding.

\begin{figure}
    \centering
\subfloat[\label{fig:hard_1}]{%
\includegraphics[height=2.5cm,width=1.\linewidth]{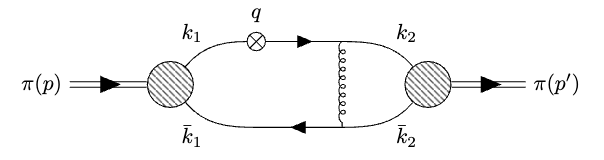}%
}\hfill
\subfloat[\label{fig:hard_2}]{%
\includegraphics[height=2.5cm,width=1.\linewidth]{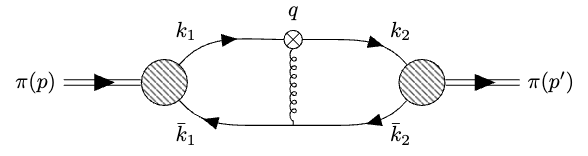}%
}\hfill
\subfloat[\label{fig:hard_3}]{%
\includegraphics[height=2.5cm,width=1.\linewidth]{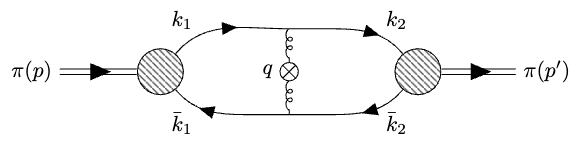}%
}
\caption{Hard contributions 
to the pion GFF~\cite{Shuryak:2020ktq,Tong:2021ctu}:  (a) quark contribution; (b) and (c)  gluon contribution. The dashed blob refers to the LFWFs, and the crossed circle to the insertion of the QCD stress tensor .}
\label{fig:hard}
\end{figure}

\section{Hard contribution to the pion gravitational formfactor}
\label{SECIII}
%\section{Pion EMT}
\label{VACUUM} 
The symmetric and conserved stress tensor (also known as the energy-momentum tensor or EMT)
is
\bea
\label{1}
T^{\mu\nu}
=-F^{a\mu\lambda}F^{a\nu}_\lambda+\frac 14 g^{\mu\nu}F^2+ \overline\psi \gamma^{(\mu} i\overleftrightarrow D^{\nu)}\psi\nonumber\\
\eea
with $\overleftrightarrow{D}_\mu=\frac{1}{2}(\overrightarrow{D}_\mu+\overleftarrow{D}_\mu)$ where $\bar{\psi}\overleftarrow{D}_\mu=-\partial_\mu\bar{\psi}+i\bar{\psi}A_\mu$ and $()$ denotes symmetrization. Here the equation of motion $(i\slashed{D}-m)\psi=0$ is already applied in~\eqref{1}. It traces to zero classically, but is anomalous quantum mechanically owing to its sensitivity to short distance fluctuations (see below).

 The quantum breaking of conformal symmetry is best captured in the channel coupled to the $trace$ of the stress 
tensor, namely
\be
\label{3X}
T^\mu{}_\mu=\frac{\beta(g^2)}{4g^4}F^a_{\mu\nu}F^{a\mu\nu}+m\overline\psi\psi
\ee
The scale dependence here comes via the running coupling constant and 
  the Gell-Mann-Low $\beta(g^2)$ function. Up to two loops
  it is
\be
\label{beta}
\beta(g^2)=-\frac{bg^4}{8\pi^2}-\frac{\bar bg^6}{2(8\pi^2)^2}+{\cal O}(g^8)
\ee

\subsection{Perturbative QCD contribution}
The leading order generic QCD perturbative contributions to the pion GFF, are illustrated in Fig.~\ref{fig:hard}. We note that in~\cite{Shuryak:2020ktq}  only the diagram  in Fig~\ref{fig:hard_1} for both  twist 2,3, was included,
while in~\cite{Tong:2021ctu,Tong:2022zax} all the diagrams with twist-2 
were considered. Here we extend the perturbative contributions in the hard region to include all contributions up-to twist-3, including the semi-hard
contribution from the instanton non-zero-modes. 

With this in mind, the complete graviton coupling in QCD can be split into three parts $$T^{\mu\nu}=\frac{\delta S_{QCD}}{\delta g_{\mu\nu}}=T_q^{\mu\nu}+T_{qg}^{\mu\nu}+T_g^{\mu\nu}$$
where
\begin{eqnarray}
 \label{VERTEX}
T_q^{\mu\nu}=&&\bar{\psi}i\gamma^{(\mu}\overleftrightarrow{\partial}^{\nu)}\psi-g^{\mu\nu}\bar{\psi}(i\overleftrightarrow{\slashed{\partial}}-m)\psi \nonumber\\
T_{qg}^{\mu\nu}=&&g_s\left[\bar{\psi}\gamma^{(\mu}A^{\nu)}\psi-g^{\mu\nu}\bar{\psi}\slashed{A}\psi\right] \nonumber\\
T_g^{\mu\nu}=&&-F^{\mu\lambda}F^{\nu}{}_{\lambda}+\frac{1}{4}g^{\mu\nu}F^2 
\end{eqnarray}
 The ensuing vertices are
\begin{widetext}
\begin{eqnarray}
T_q^{\mu\nu}&&\rightarrow\left(
\begin{tikzpicture}[scale=0.6,baseline=(o)]
   \begin{feynhand}
   \path (0,0) -- (4,0);
    \vertex  (a) at (-0.5,0);   
    \vertex[crossdot] (b) at (2,0) {}; 
    \vertex  (c) at (4.5,0);
    \vertex (o) at (5,-0.1);
    \node at (0.75, 0.5) {$p_1$};
    \node at (3.25,0.5) {$p_2$};
    \propag [fer] (a) to (b);
    \propag [fer] (b) to (c);
   \end{feynhand}
\end{tikzpicture}\right)
=\frac{1}{2}\gamma^{(\mu}(p_1+p_2)^{\nu)}-\frac{1}{2}g^{\mu\nu}(\slashed{p}_1+\slashed{p}_2)\nonumber\\
T_{qg}^{\mu\nu}&&\rightarrow\left(\begin{tikzpicture}[scale=0.6,baseline=(o)]
   \begin{feynhand}
   \path (0,0) -- (4,0);
    \vertex  (a) at (-0.5,0);   
    \vertex[crossdot] (b) at (2,0) {}; 
    \vertex  (c) at (4.5,0);
    \vertex (d) at (2,-1.5){$\rho$};
    \vertex (o) at (5,-0.1);
    \node at (0.75, 0.5) {$p_1$};
    \node at (3.25,0.5) {$p_2$};
    \propag [fer] (a) to (b);
    \propag [fer] (b) to (c);
    \propag [glu] (b) to (d);
   \end{feynhand}
\end{tikzpicture}\right)
=g_s\left[\gamma^{(\mu}g^{\nu)\rho}-g^{\mu\nu}\gamma^{\rho}\right]\nonumber\\
T_g^{\mu\nu}&&\rightarrow\left(\begin{tikzpicture}[scale=0.6,baseline=(o)]
   \begin{feynhand}
   \path (0,0) -- (4,0);
    \vertex  (a) at (-0.5,0){$\rho$};   
    \vertex[crossdot] (b) at (2,0) {}; 
    \vertex  (c) at (4.5,0){$\lambda$};
    \vertex (o) at (5,-0.1);
    \node at (0.75, 0.5) {$p_1$};
    \node at (3.25,0.5) {$p_2$};
    \propag [glu] (a) to (b);
    \propag [glu] (b) to (c);
   \end{feynhand}
\end{tikzpicture}\right)
=g_{\rho\lambda}p_1^\mu p_2^\nu-g^{\mu\rho}p_1^\lambda p_2^{\nu}-g^{\nu\rho}p_1^\lambda p_2^{\mu}+ p_1\cdot p_2 g^{\mu\nu}g^{\rho\lambda}+(1\leftrightarrow2)\nonumber\\
\end{eqnarray}

\end{widetext}
 After including the perturbative gluonic contribution, the hard contribution to the scalar and mass form factors take the form of two wave functions convoluted with the hard vertex
\begin{widetext}
 \begin{eqnarray}
 \label{eqn_T00PT}
T^\pi_{00,\mathrm{pQCD}}(Q^2)=&&\bigg(\frac{4\pi\alpha_sC_Ff_\pi^2}{N_c}\bigg)\int_0^1 dx_1\int_0^1dx_2\,
\left(\frac {x_1}{\bar x_1\bar x_2}+\frac{1}{2x_1\bar{x}_1}\right)\varphi^A_\pi(x_1)\varphi^A_\pi(x_2)\nonumber\\
&& +\frac{2\chi_\pi^2}{Q^2}\left[\left(\frac {1}{\bar x_1\bar x_2}\left(\frac12-x_1\right)+\frac{1}{4x_1\bar{x}_2}-\frac {1}{2 x_1\bar x_1}\right)\varphi_\pi^P(x_1)\varphi_\pi^P(x_2)
+\frac {x_1}{\bar x_1\bar x_2}\varphi_\pi^P(x_2)\frac{\varphi_\pi^{T\prime}(x_1)}6\right]\nonumber\\
T^\pi_{\mu\mu,\mathrm{pQCD}}(Q^2)=&&-3\bigg(\frac{4\pi\alpha_sC_Ff_\pi^2}{N_c}\bigg)\frac{2\chi_\pi^2}{Q^2}\int_0^1 dx_1\int_0^1dx_2\ \frac{1}{\bar{x}_1\bar{x}_2}\left(\frac 1{\bar x_1}-1\right)\left[\varphi_\pi^P(x_1)\varphi_\pi^P(x_2)
+\varphi_\pi^P(x_2)\frac{\varphi_\pi^{T\prime}(x_1)}6\right]\nonumber\\
 \end{eqnarray}
\end{widetext}
The results (\ref{eqn_T00PT}) for  $T^\pi_{\mu\mu}$ are  in agreement with those quoted in~\cite{Tong:2021ctu} at the twist-2 level. Our results include the twist-3 contributions.

We now observe that the $T^\pi_{\mu\mu}$ constructed by the $A_{q,g}$ and $D_{q,g}$ in~\cite{Tong:2021ctu} is also zero at twist-2. This is consistent  with the fact that the scalar operator is a twist-3 operator. Interestingly, the hard contribution to the gluonic scalar form factor obtained from the vertex in \ref{fig:hard_3} has a non-zero contribution at the leading twist. To this order, the perturbative calculations does not capture the trace anomaly. Therefore, we will treat the trace anomaly as a completely  non-perturbative effect, 
\begin{widetext}
\begin{equation}
\begin{aligned}
     &G_{\pi,\mathrm{pQCD}}(Q^2)\\
     =&-\frac{\alpha_sb}{16\pi m^2_\pi}\left[\frac{4\pi\alpha_sC_F}{N_c}f_\pi^2\int_0^1dx_1\int_0^1dx_2\left(\frac{1}{x_1\bar{x}_2}+\frac{1}{x_2\bar{x}_1}\right)\left(\varphi^A_\pi(x_1)\varphi^A_\pi(x_2)-\frac{3\chi_\pi^2}{Q^2}\varphi^P_\pi(x_1)\varphi^P_\pi(x_2)\right)\right]
\end{aligned}
\end{equation}
\end{widetext}
where $b=\frac{11}{3}N_c-\frac 23 N_f$.
This result is consistent with the one in~\cite{Tong:2022zax}, and is now extended to include the twist-3 contribution. Finally, we note that the trace-part of the pion EMT does not exhibit a power decay asymptotically. It asymptotes a constant
a constant modulo logarithms from $\alpha_s$ at large $Q^2$ \cite{Tong:2021ctu}.

The {\it apparent} separation of the quark and gluon contributions $A_{q,g}$ and $D_{q,g}$ in the pion EMT, follows from enforcing the invariant decomposition
\bea
\label{EMTqg}
&&\left<p'\left|T_{q,g}^{\mu\nu}\right|p\right>=2 {\bar{p}^\mu \bar{p}^\nu}\,A_{q,g}(q)\nonumber\\
&&+
\frac 12 ({q^\mu q^\nu}-g^{\mu\nu}q^2)\,D_{q,g}(q)+2m^2_\pi g^{\mu\nu}\bar{C}_{q,g}(q)
\nonumber\\
\eea
The ensuing quark contributions from Fig.~\ref{fig:hard_1},\ref{fig:hard_2}, are
\begin{widetext}
\begin{eqnarray}
    A_q(Q^2)=&&\bigg(\frac{4\pi\alpha_sC_Ff_\pi^2}{N_cQ^2}\bigg)\int_0^1 dx_1\int_0^1dx_2\,
\frac {1}{\bar x_1\bar x_2}\bigg\{(x_1+x_2+1)\varphi^A_\pi(x_1)\varphi^A_\pi(x_2)\nonumber\\
&& +\frac{2\chi_\pi^2}{Q^2}\left[2\left(\frac 1{\bar x_1}-x_1-1\right)\varphi_\pi^P(x_1)\varphi_\pi^P(x_2)
+3\left(\frac 1{\bar x_1}-\bar x_1\right)\varphi_\pi^P(x_2)\frac{\varphi_\pi^{T\prime}(x_1)}6\right]\bigg\}\nonumber\\
D_q(Q^2)=&&\bigg(\frac{4\pi\alpha_sC_Ff_\pi^2}{N_cQ^2}\bigg)\int_0^1 dx_1\int_0^1dx_2\,
\frac {1}{\bar x_1\bar x_2}\bigg\{(x_1+x_2-3)\varphi^A_\pi(x_1)\varphi^A_\pi(x_2)\nonumber\\
&&+\frac{2\chi_\pi^2}{Q^2}\left[\left(2+2\bar x_1\right)\varphi_\pi^P(x_1)\varphi_\pi^P(x_2)
- \left(\frac3 {\bar x_1}-3+x_1 \right)\varphi_\pi^P(x_2)\frac{\varphi_\pi^{T\prime}(x_1)}6\right]\bigg\}\nonumber\\
\bar{C}_q(Q^2)=&&\frac{1}{4m_\pi^2}\bigg(\frac{4\pi\alpha_sC_Ff_\pi^2}{N_c}\bigg)\int_0^1 dx_1\int_0^1dx_2\,
\frac {2}{\bar x_1\bar x_2}\Bigg\{\bar{x}_1\varphi^A_\pi(x_1)\varphi^A_\pi(x_2)\nonumber\\
&&-\frac{2\chi_\pi^2}{Q^2}
\left(\frac{1}{\bar x_1}- x_1+\frac 12\right)\varphi_\pi^P(x_1)\varphi_\pi^P(x_2)
\Bigg\}
\end{eqnarray}
%\end{widetext}
while the gluon contributions from Fig.\ref{fig:hard_3}, read
%\begin{widetext}
\begin{eqnarray}
    A_g(Q^2)=&&\bigg(\frac{4\pi\alpha_sC_Ff_\pi^2}{N_cQ^2}\bigg)\int_0^1 dx_1\int_0^1dx_2\,
\bigg\{\frac {1}{x_1\bar x_1}\varphi^A_\pi(x_1)\varphi^A_\pi(x_2)\nonumber\\
&& +\frac{2\chi_\pi^2}{Q^2}\left[\bigg(\frac{3}{4x_1\bar x_2}-\frac 3{2x_1\bar x_1}+\frac{1}{\bar x_1 \bar x_2}\left(\frac{1}{\bar x_1}-x_1+\frac 12\right)\bigg)\varphi_\pi^P(x_1)\varphi_\pi^P(x_2)
\right]\bigg\}\nonumber\\
D_g(Q^2)=&&\bigg(\frac{4\pi\alpha_sC_Ff_\pi^2}{N_cQ^2}\bigg)\int_0^1 dx_1\int_0^1dx_2\,
\bigg\{\frac {1}{x_1\bar x_1}\varphi^A_\pi(x_1)\varphi^A_\pi(x_2)\nonumber\\
&&+\frac{2\chi_\pi^2}{Q^2}\left[\left(\frac{1}{2x_1\bar x_1}-\frac1{4x_1\bar x_2}-\frac3{\bar x_1\bar x_2}\left(\frac{1}{\bar x_1}-x_1+\frac12\right)\right)\varphi_\pi^P(x_1)\varphi_\pi^P(x_2)
\right]\bigg\}\nonumber\\
\bar{C}_g(Q^2)=&&-\bar{C}_q(Q^2)
\end{eqnarray}
\end{widetext}
At the twist-2 level, $\bar{C}_q(Q^2)=Q^2D_g(Q^2)/(4m^2_\pi)$ which is consistent with the result in~\cite{Tong:2021ctu}. Our results show that this observation does not extend to twist-3.

\subsection{Semi-hard instanton non-zero mode contribution}
%\begin{figure}
%    \centering
%    \includegraphics[height=6cm,width=.99\linewidth]{G_Q.png}
%\caption{Hard gluon contribution to the trace of the pion EMT from the ILM (orange-solid line), from the hard pQCD (green-solid line),  and the recent lattice results~\cite{Hackett:2023nkr}.}
%\label{fig:G_Q}
%\end{figure}
In leading order in the instanton packing fraction $\kappa=\pi^2\rho^4n_{I+A}$, only the  non-zero modes illustrated in Fig.~\ref{fig:hard}c contribute to the pion EMT
(the zero modes contribute through molecules at next-to-leading order in $\kappa$ as we discussed earlier). The results including the twist-2,3 pion DAs, are ~\cite{Shuryak:2020ktq}

\begin{widetext}
  \begin{eqnarray}
 \label{eqn_T00NZM}
T^\pi_{00,\mathrm{NZM}}(Q^2)=&&\left(\frac{\kappa\pi^2f_\pi^2\chi_\pi^2}{N_cM^2}\right)\rho Q~\mathbb G_V(\rho Q)\int dx_1dx_2\,(\bar x_1-\bar x_2)\nonumber\\
&&\times\bigg(\bar x_1\varphi_\pi^P(x_1)\varphi_\pi^P(x_2)+(\bar x_2-\bar x_1)\varphi_\pi^P(x_2)\frac{\varphi_\pi^{T\prime}(x_1)}6
-\bar x_1\frac{\varphi_\pi^{T\prime}(x_1)}6\frac{\varphi_\pi^{T\prime}(x_2)}6\bigg)\nonumber\\
T^\pi_{\mu\mu ,\mathrm{NZM}}(Q^2)=&&\left(\frac{2\kappa\pi^2f_\pi^2\chi_\pi^2}{N_cM^2}\right)\rho Q~\mathbb G_V(\rho Q)\int dx_1dx_2\,\bar x_1 x_2\nonumber\\
&&\times\bigg(\varphi_\pi^P(x_1)\varphi_\pi^P(x_2)-\varphi_\pi^P(x_2)\frac{\varphi_\pi^{T\prime}(x_1)}6
-\frac{\varphi_\pi^{T\prime}(x_1)}6\frac{\varphi_\pi^{T\prime}(x_2)}6\bigg)
 \end{eqnarray}
The induced vector form factor ${\mathbb G}_V$  is
\begin{equation}{\mathbb G}_V(Q\rho)=F_1(Q\rho)+{1 \over N_c M^2\rho^2}F_2(Q\rho)
\label{eqn_GV}\end{equation}
with 
\begin{eqnarray}
F_1(x)\equiv  \bigg(\frac {K_1(x)}{x}\bigg)^{\prime\prime}=&&
{1 \over 4 x^3}(4 x K_0(x) + (8 + 3 x^2) K_1( x) + 
x (4 K_2( x) + x K_3(x)))\nonumber\\
F_2(x)\equiv x \bigg(\frac{(x K_1)^\prime}{x}\bigg)^\prime=&&
{1 \over 4 x}(-2 x K_0(x) + (-4 + 3 x^2) K_1(x) + 
x (-2 K_2(x) + x K_3(x)))\equiv xK_1(x)\nonumber\\
\label{eqn_F2}
\end{eqnarray}
\end{widetext}

\begin{figure}
    \centering
\includegraphics[height=2.5cm,width=1.\linewidth]{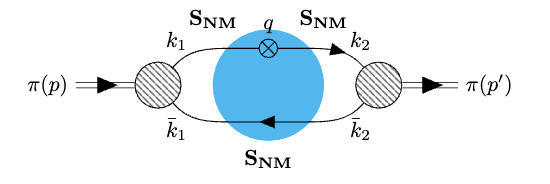}
\caption{Semi-hard contributions to the pion GFF~\cite{Shuryak:2020ktq}. The blue insertion refers to the pseudoparticle (I+A). The labels $\rm S_{NM}$ refer to  the distorted non-zero-mode propagators. }
\label{fig:Non_zero}
\end{figure}

%\begin{widetext}
\begin{figure*}
%[ht!]
\centering
\subfloat[\label{gluonHallDataX}]{%
\includegraphics[height=5.15cm,width=.485\linewidth]{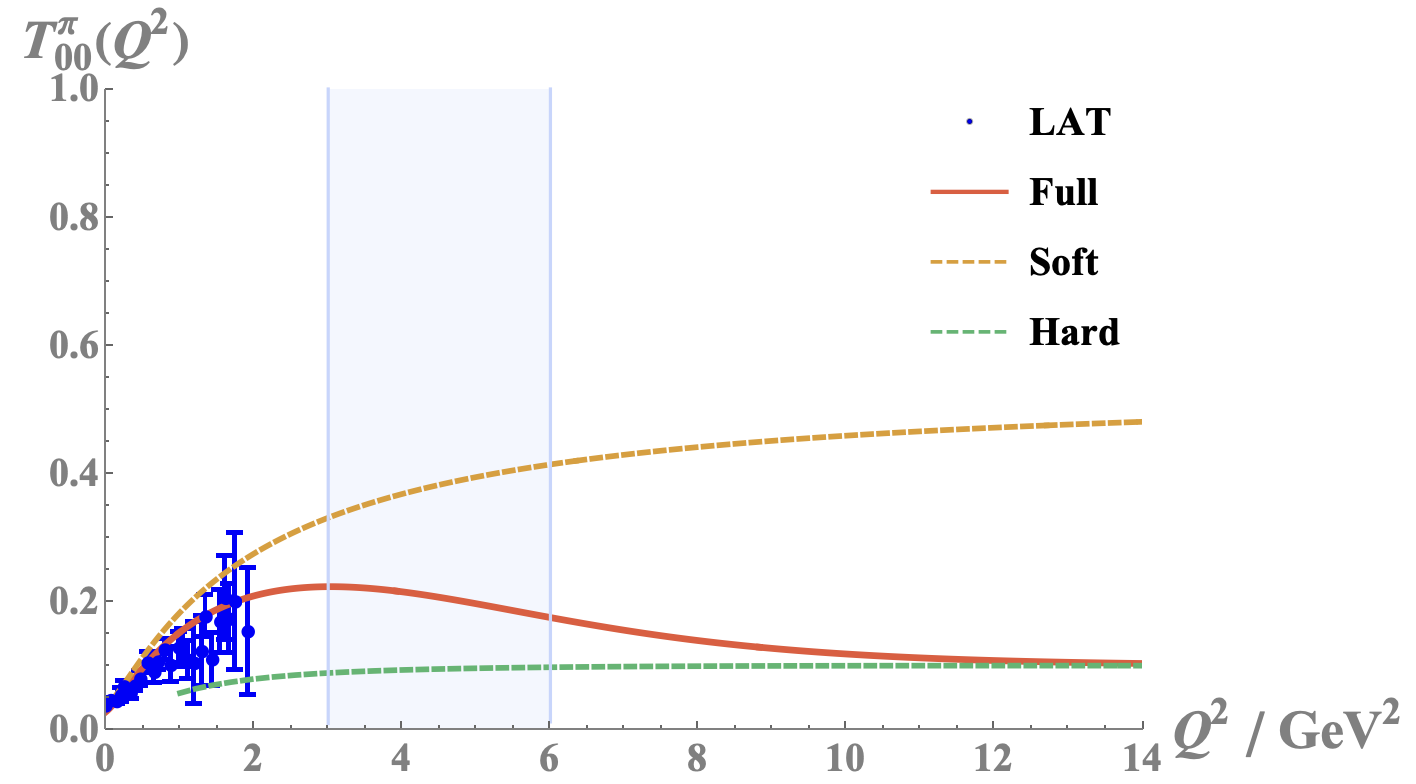}%
}\hfill
%\subfloat[\label{seaHallData}]{%
%\includegraphics[height=5cm,width=.45\linewidth]{T00_all_Q_2.png}%
%}\hfill
\subfloat[\label{uValenceHallDataX}]{%
\includegraphics[height=5.15cm,width=.485\linewidth]{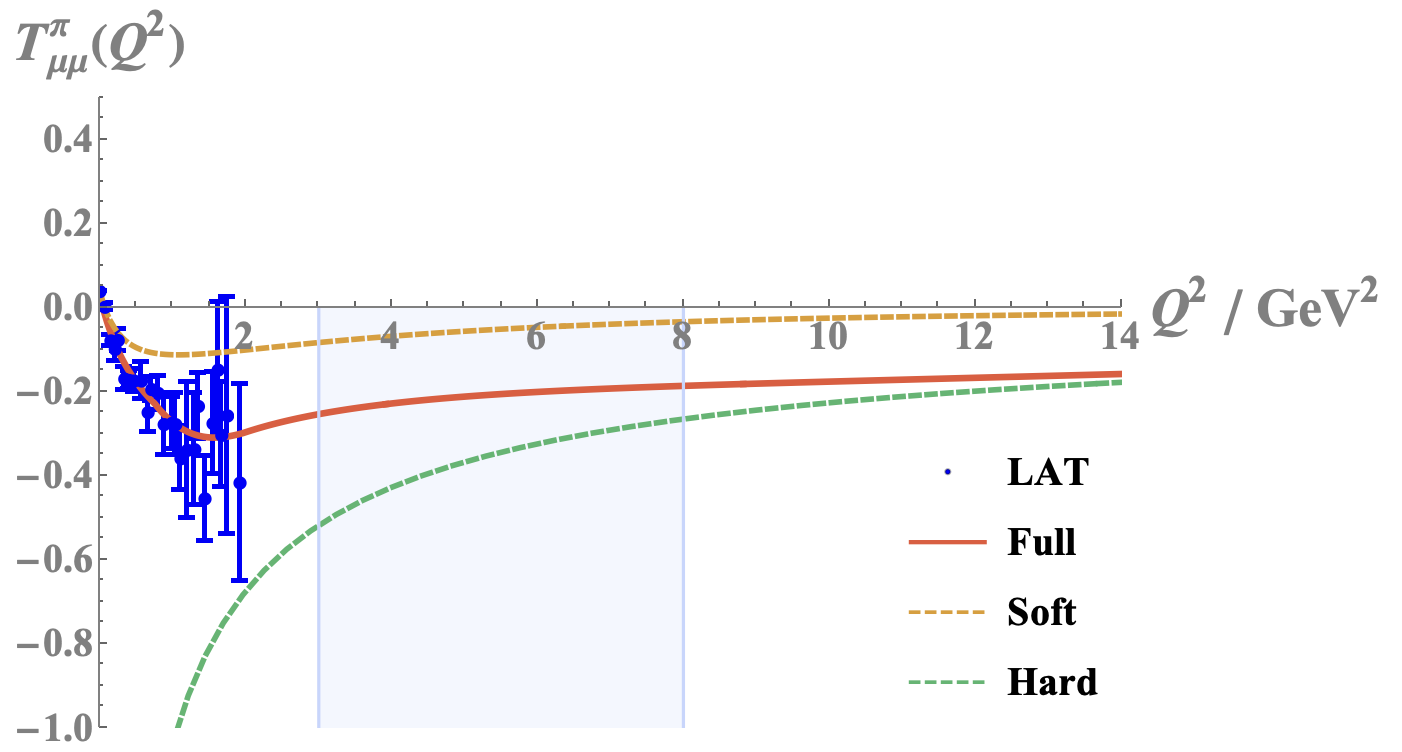}%
}
\caption{The pion 00-EMT (a) and pion trace-EMT (b) versus $Q^2$, with the  soft contribution from the ILM (dashed-orange line) from~\cite{Liu:2024jno},
the hard plus semi-hard contribution (dashed-green line) given in ~\eqref{eqn_T00PT}-\eqref{eqn_T00NZM},
and the interpolation full sum (solid-red line) in \eqref{INTER}, compared to the lattice results from~\cite{Hackett:2023nkr}.}
%The red curve is done by using the linear superposition of the soft curve (yellow) and hard curve (green). $c_{\mathrm{soft}}A_\pi^{\mathrm{soft}}+c_{\mathrm{hard}}A_\pi^{\mathrm{hard}}$ such that on the left side of the band $(c_{\mathrm{soft}},c_{\mathrm{hard}})=(1,0)$ and on the right side of the band $(c_{\mathrm{soft}},c_{\mathrm{hard}})=(0,1)$ (a) The band is $3$ - $6$ GeV$^2$ (b) The band is $3$ - $8$ GeV$^2$. The comparison is to the recent lattice results in~\cite{Hackett:2023nkr}.}
\label{fig:T00TMUMU}
\end{figure*}
%\end{widetext}

\begin{figure*}
%[ht!]
\centering
\subfloat[\label{uValenceHallData}]{%
\includegraphics[height=5.15cm,width=.485\linewidth]{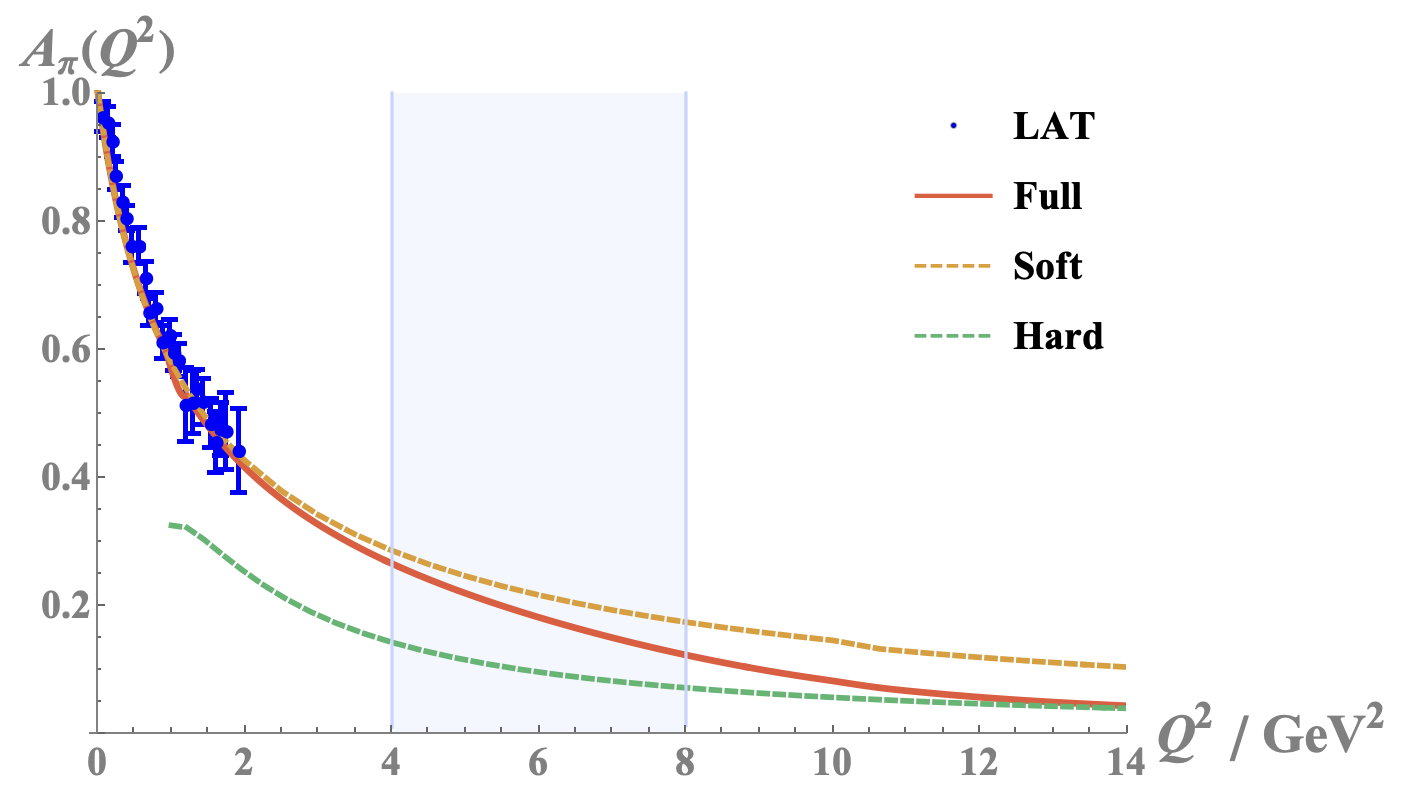}%
}\hfill
\subfloat[\label{dValenceHallData}]{%
\includegraphics[height=5.15cm,width=.485\linewidth]{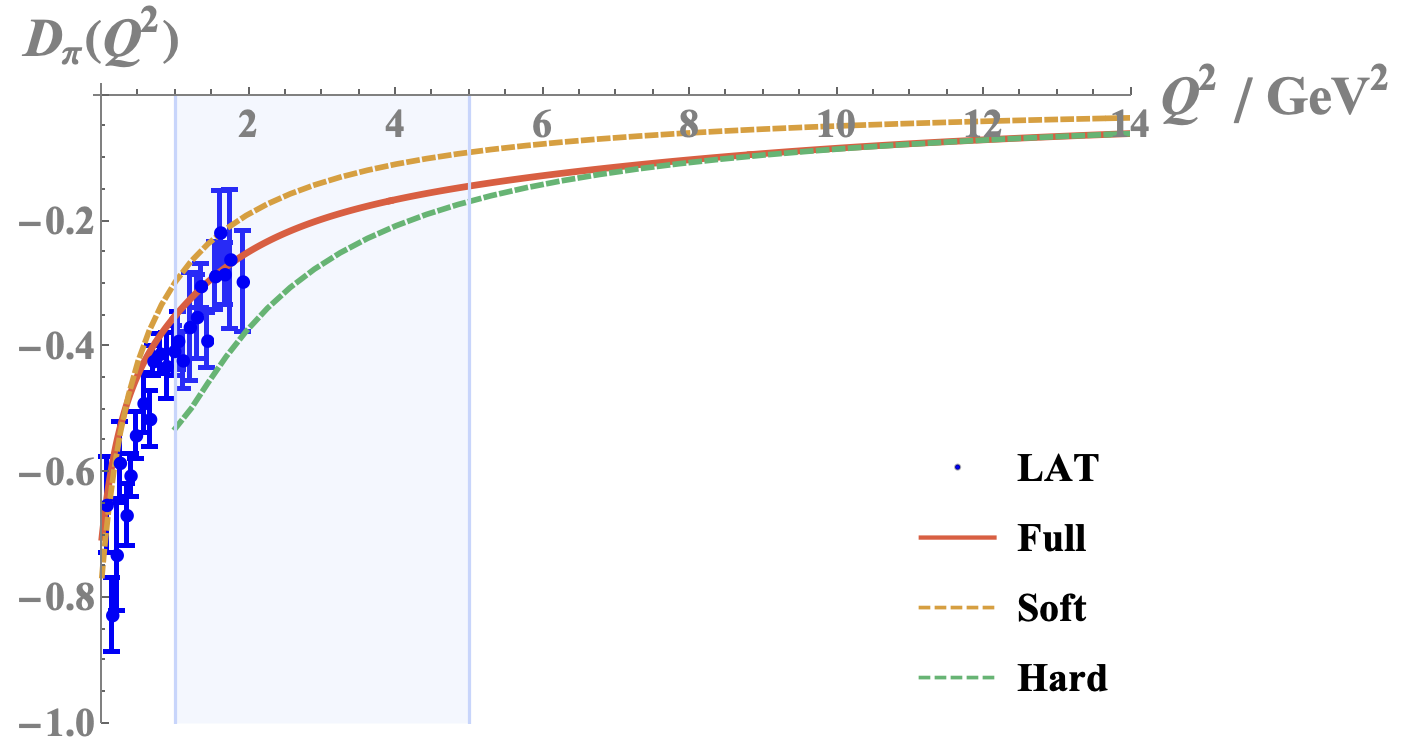}%
}
\caption{The pion A-GFF (a) and pion D-GFF (b) versus $Q^2$ given in~\eqref{AQDQX}, with the  soft contribution from the ILM (dashed-orange line) from~\cite{Liu:2024jno},
the hard plus semi-hard contribution (dashed-green line) given in ~\eqref{eqn_T00PT}-\eqref{eqn_T00NZM}, and the interpolation full sum (solid-red line) in \eqref{INTER}, compared to the lattice results from~\cite{Hackett:2023nkr}.}
%The calculated formfactors $A_\pi(Q^2)$ and $D_\pi(Q^2)$ (lines) compared to the recent lattice results of \cite{Hackett:2023nkr} (points).
%The red curves are  linear superposition of the ``soft" curve (yellow) and ``hard" curve (green), $c_{\mathrm{soft}}A_\pi^{\mathrm{soft}}+c_{\mathrm{hard}}A_\pi^{\mathrm{hard}}$. such that on the left sides of the band (shown by light blue) ($(c_{\mathrm{soft}},c_{\mathrm{hard}})=(1,0)$ and on the right side of the band $(c_{\mathrm{soft}},c_{\mathrm{hard}})=(0,1)$. The band in (a)  is $4$ - $8$ GeV$^2$  and in (b) it  is $1$ - $5$ GeV$^2$.  }
\label{fig:ADALL}
\end{figure*}

%\begin{widetext}
\begin{figure*}
%[ht!]
\centering
\subfloat[\label{gluonHallData}]{%
\includegraphics[height=5.1cm,width=.48\linewidth]{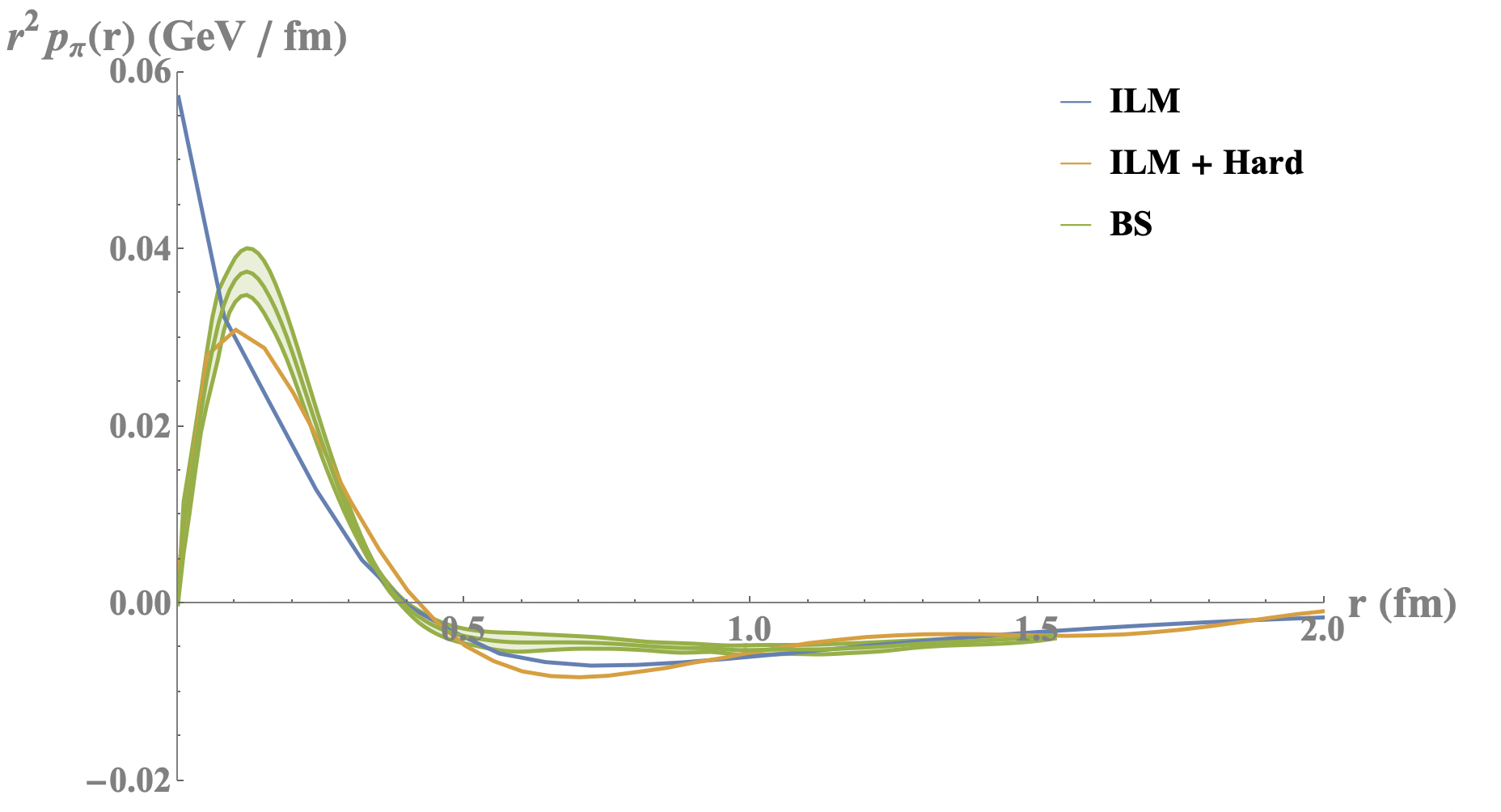}%
}\hfill
\subfloat[\label{seaHallData}]{%
\includegraphics[height=5.1cm,width=.47\linewidth]{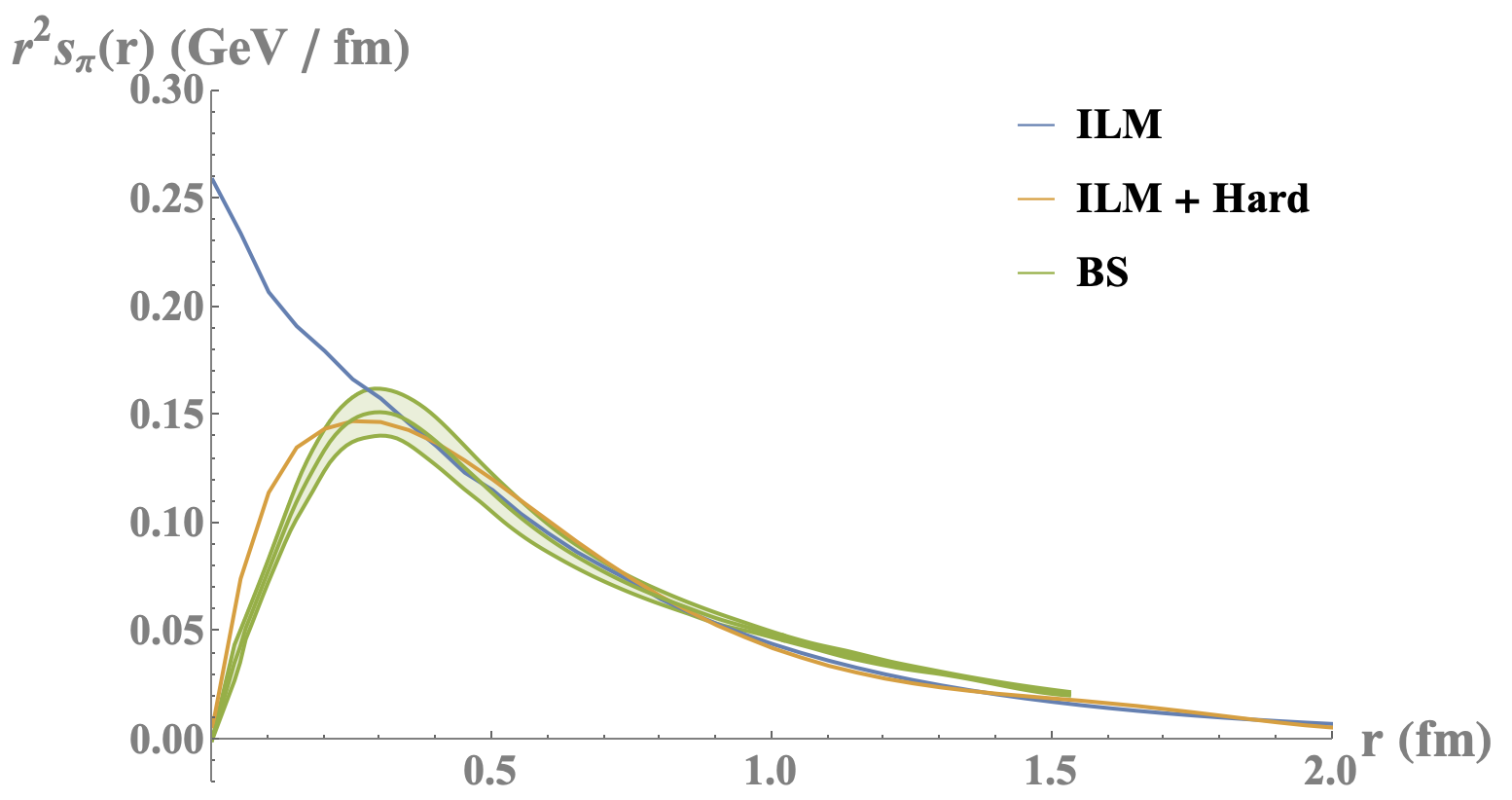}%
}
\caption{The pressure (a) and shear (b) in the pion calculated using the ILM (blue-solid line) and the ILM plus pQCD (orange solid line). The comparison is to the recent results from the
Bethe-Salpeter resummation~\cite{Xu:2023izo}.}
\label{fig:PS}
\end{figure*}
%\end{widetext}

\section{Pion GFFs}
\label{SECIV}
\label{RESULTS}
The pion 00-EMT and trace-EMT can be recast  in terms of the pion GFFs,  
\begin{widetext}
\bea
\label{MOM2X}
T^\pi_{00}(Q^2)=\left<p'|T^{00}|p\right>&=&
2 \bigg(m_\pi^2+\frac 14 Q^2\bigg)A_\pi(Q^2)+\frac 12 Q^2D_\pi(Q^2)\nonumber\\
%\left<p_2|T^{-+}|p_1\right>&=&
%\bigg(m_\pi^2+\frac 14 Q^2\bigg)A(Q^2)+\frac 12 Q^2D(Q^2)\nonumber\\
T^\pi_{\mu\mu}(Q^2)=\left<p'|T^\mu{}_\mu|p\right>&=&
2 \bigg(m_\pi^2+\frac 14 Q^2\bigg)A_\pi(Q^2)+\frac 32 Q^2D_\pi(Q^2)
\eea
or equivalently 
%\begin{widetext}
\bea
\label{AQDQX}
A_\pi(Q^2)&=&\frac 1{Q^2+4m_\pi^2}
\bigg(3 
\left< p'|T^{00}|p\right>-
\left< p'|T^{\mu}_\mu|p\right>\bigg)
\nonumber\\
D_\pi(Q^2)&=&
\frac {1}{Q^2}\bigg(\left<p'|T^\mu_\mu|p\right>
-\left<p'|T^{00}|p\right>\bigg)
\eea
\end{widetext}

To analyze the contributions stemming from the ILM, we use the  standard ILM parameters we noted earlier, with a mean vacuum EMT
\be
\label{SCALE}
\langle T^\mu{}_\mu\rangle\approx -
\frac {b\langle F^2\rangle}{32\pi^2}
\approx-b~n_{I+A}
\approx -10~{\rm fm}^{-4}
\ee
with $b\simeq\frac {11}3N_c$ in large $N_c$ limit, a packing fraction $\kappa\approx \pi^2\rho^4n_{I+A}\approx 0.1$,
 a mean instanton size $\rho=0.313\,\rm fm$, a quark constituent mass
$M=398\,\rm MeV$  and  a pion mass $m_\pi=135\,\rm MeV$. Also, for the hard contributions, we will make use of the gluon and the quark transverse energies    $m_{\rm gluon}^2=1\,\rm GeV^2$ and $E_\perp^2=0.3\,\rm GeV^2$ estimated in~\cite{Shuryak:2020ktq},  to which we refer for further details.

In  Fig.~\ref{gluonHallDataX} we show  our final results for the pion 00-EMT. The twits 2,3 hard including the semi-hard instanton contribution are shown in dashed-green, the soft
contributions from our recent analysis~\cite{Liu:2024jno}, are shown in dashed-brown. The full
contribution in solid-red interpolates between the soft and hard contributions by optimal matching in the blue-band, 
 the coefficients $c_{\rm soft, hard}$ in the linear combination
\bea
\label{INTER}
c_{\mathrm{soft}}A_\pi^{\mathrm{soft}}+c_{\mathrm{hard}}A_\pi^{\mathrm{hard}}
\eea
The band width is fixed to $3$ - $6$ GeV$^2$, the left side of the band $(c_{\mathrm{soft}},c_{\mathrm{hard}})\rightarrow(1,0)$ and on the right side of the band $(c_{\mathrm{soft}},c_{\mathrm{hard}})\rightarrow(0,1)$. 
In Fig.~\ref{uValenceHallDataX} we show our final results for the pion trace-EMT, with the color coding following that in Fig.~\ref{gluonHallDataX}, with the  matching band width fixed to $3$ - $8$ GeV$^2$. Our final results in solid-red, interpolating between the soft and hard regions are compares well in the soft regime, to the recent lattice results blue data from~\cite{Hackett:2023nkr}.

In Fig.~\ref{uValenceHallData} we show our results for the pion A-form factor $A_\pi(Q)$ versus $Q^2$. Again, the pQCD twits 2,3 hard plus the instanton semi-hard contribution, are shown in dashed-green, the soft
contributions from~\cite{Liu:2024jno} are shown in dashed-brown. The full
contribution in solid-red interpolates between the soft and hard contributions by optimal matching in the blue-band, 
 the coefficients $c_{\rm soft, hard}$ as detailed above.
 In Fig.~\ref{dValenceHallData} we show our results for the pion D-form factor $D_\pi(Q)$ versus $Q^2$, with the same color labeling. Our final interpolating result in solid-red is compared to the recent lattice data in blue from~\cite{Hackett:2023nkr}. Again, the soft contribution from the ILM agrees well with the lattice results.

\section{Pressure and shear inside the pion}
\label{SECV}
The pion $D$ form factor allows for  the pion pressure and shear force distributions. In the Breit frame they are defined as~\cite{Polyakov:2018zvc}
\begin{widetext}
\bea
\label{PRESSURE}
p_\pi(r)&=&\frac 1{6r^2}\frac{d}{dr}r^2\frac{d}{dr}\left[\frac{1}{4\pi^2r}\int_0^\infty dQ^2\frac{D_\pi(Q^2)\sin(Qr)}{E_\pi(Q)}\right]
=-\frac 1{6\pi^2 r}\int_0^\infty dQ\, \frac {Q^3{\rm sin}(Qr)}{2E_\pi(Q)} \,D_\pi(Q^2)
\nonumber\\
s_\pi(r)&=&-\frac 3{8}r\frac{d}{dr}\frac{1}{r}\frac{d}{dr}\left[\frac{1}{4\pi^2r}\int_0^\infty dQ^2\frac{D_\pi(Q^2)\sin(Qr)}{E_\pi(Q)}\right] 
=-\frac 3{8\pi^2}\int_0^\infty dQ\,
\frac{Q^4j_2(Qr)}{2E_\pi(Q)}\,D_\pi(Q^2)
\nonumber\\
\eea
\end{widetext}
with $E^2_\pi(Q)=m_\pi^2+\frac 14 Q^2$.  Note that our  pion $A$,$D$-form factors were derived in the soft and semi-hard regimes, and matched to the hard regime
for the whole range of distances, as we detailed above. We have checked that the our pion pressure satisfies the ``stability constraint" \cite{Polyakov:2018zvc}
\begin{equation}
    \int_0^\infty dr 4\pi r^2 p_\pi(r)=0
\end{equation}
and that the pion  shear  satisfies the D-term sum rule
\begin{equation}
    -\frac{32\pi }{45}m_\pi\int_0^\infty dr r^4 s_\pi(r)=D_\pi(0)
\end{equation}

In Fig.~\ref{fig:PS}a we show our results for the pion radial pressure from the ILM (solid-blue) and the ILM+hard contribution after matching (solid-orange), versus the radial distance. The
comparison is to the recent results using the Bethe-Salpeter 
result (green-band) from~\cite{Xu:2023izo}. In Fig.~\ref{fig:PS}b we show our results for the pion shear,
with the same color coding and comparison as in Fig.~\ref{fig:PS}a. Our results show clearly the range of importance of the soft and hard contributions as they cross at about $0.5\,\rm fm$.

\section{Conclusion}
\label{SECVI}
In QCD, the two main nonperturbative phenomena, i.e. the breaking of
$SU(N_f)$ chiral symmetry and the breaking of
conformal symmetry, are both related to the famed ``anomaly relations". The relationship of the axial anomaly to the pions has been  known since 
1960's. However, the relationship of the ``scale anomaly" 
is less known, but yields stringent constraints on the bulk hadronic correlations in the form of low energy theorems~\cite{Novikov:1981xi}. 

Both phenomena are at work 
 in the QCD instanton vacuum models \cite{Schafer:1996wv}. The scale anomaly constraints take the form of non-Poisson fluctuations in the number of pseudo particles. The ``vacuum compressibility" 
of the instanton ensemble is indicative of a quantum liquid. 

While these phenomena in the QCD vacuum are well known, 
their quantitative consequences for specific hadrons have
not yet been worked out in detail solely from the vacuum. In this paper, we derived
the gravitational  form factors of the pion, focusing on 
the important semi-hard region, located between the ``soft" (domain of chiral
models) and ``hard" (perturbative) regions. We have derived
the instanton-induced contributions for the form factors in semi-hard region, and shown that match well with the hard region, for which the perturbative predictions
for the EMT were only addressed recently. 
Our results can be tested by future lattice and, of course, experimental measurements.

Let us start with the pion mass. About half of it is due to the
breaking of conformal symmetry in the QCD instanton vacuum through the trace anomaly, while the other half stems
 from the scalar 
sigma term. 
This observation extends to the trace part of the pion EMT, with a short distance contribution following from the scalar glueballs, and a long distance contribution stemming from the pion scalar form factor,
which is mostly a 2-pion cloud. The gluon mass radius of the pion, is about a-third that of the scalar pion cloud. 

We also provided the explicit forms for the light front twist 2,3 pion wave functions. Modulo kinematics, these wave functions are profiling of the fermionic zero modes. The leading light front pion EMT to the graviton, was also explicitly derived using the leading twist pion 
light front wave function. The combined result with the trace part of the pion EMT form factors, yield a complete characterization of the two invariant EMT pion form factors with spin-2 (tensor) and spin-0 (scalar).

Our results compare well with  the recent but limited lattice simulations~\cite{Hackett:2023nkr,wang2024trace}.  An extension of the lattice simulations to the hard region would be welcome, for a further test of our derivation. Our pion GFFs yield a specific characterization of hard and soft contributions to the pion radial pressure and shear, in overall agreement with the recent results using the Bethe-Salpeter approach~\cite{Xu:2023izo}.

The relevance of the pion GFFs in diffractive photoproduction on pions is yet to be explored empirically. A recent suggestion using diffractive heavy meson production on light nuclei, points at the
possibility of their empirical extraction from exchange contributions~\cite{He:2024vzz}.

\vskip 0.5cm
{\noindent\bf Acknowledgements}

\noindent 
This work is supported by the Office of Science, U.S. Department of Energy under Contract  No. DE-FG-88ER40388.
This research is also supported in part within the framework of the Quark-Gluon Tomography (QGT) Topical Collaboration, under contract no. DE-SC0023646.

\bibliography{PION,PION1}

\end{document}